\documentstyle[seceq,epsf]{ptptex}

\title{
Dynamical Properties of Euclidean Solutions \\
in a Multidimensional Cosmological Model
}

\author{
Hirotaka {\sc  Ochiai}\footnote{E-mail: ochiai@utap.phys.s.u-tokyo.ac.jp} 
and Katsuhiko {\sc Sato}\footnote{E-mail: sato@utap.phys.s.u-tokyo.ac.jp}
}

\inst{
$^{1)}$Department of Physics, School of Science, The University of Tokyo,
Tokyo 113-0033
\\
$^{2)}$Research Center for the Early Universe, School of Science, 
The University of Tokyo, Tokyo 113-0033}
\recdate{
}
\abst{
In the framework of the Hartle-Hawking no-boundary proposal,
we investigated quantum creation of the multidimensional universe
with a cosmological constant ($\Lambda$) but without matter fields.
We have found that the classical solutions of the Euclidean Einstein
equations
in this model have ``quasi-attractors'', i.e., most trajectories
on the $a$-$b$ plane,
where $a$ and $b$ are the scale factors of external and internal
spaces, go around
a point. It is presumed that  the wave function of the universe has a
hump near this quasi-attractor point. 
In the case that both the curvatures of external and internal spaces are
positive, 
and $\Lambda>0$,
there exist Lorentzian  solutions which start near the
quasi-attractor,
the internal space remains microscopic, and the external space evolves
into our macroscopic universe.
}
\begin{document}
\maketitle
\section{Introduction}
\label{sec:introduction}
In modern theories of unified physical interactions,  
spacetime has more than four dimensions. 
It is wellknown that Kaluza-Klein theory~\cite{kaluza,klein}
created interest in investigations in spacetime 
which has more than four dimensions.
Superstring theories (e.g. Ref.~\cite{polchinski} )
are at the moment the most promising candidates
for a unified description of the basic physical interactions.
There are five anomaly-free, perturbative superstring theories.
The critical dimensions of spacetime are ten for these theories.
There is now evidence that the five superstring theories are related 
by duality symmetries.
Furthermore, they are related to $N=1$, $D=11$ supergravity theory.
It is conjectured that these theories are the limits of one theory, M-theory, 
in which spacetime has eleven dimensions.

Since spacetime has four large dimensions
and some additional number of small and highly curved
spatial dimensions,
we can see only the four large dimensions.
How does compactification of internal space take place?
The first answer is through multidimensional cosmology. 
In general relativity, the geometry of spacetime is dynamical.
The three-dimensional space we observe was once as small 
as the internal space, and expanded during evolution of the universe,
while internal space contracted or has remained small 
during evolution of the universe.
Therefore, internal space is microscopic and is not observable.
This explanation is called dynamical compactification.
The second explanation involves by spontaneous symmetry breaking.
Though spacetime would have $SO$($D$$-$1,1) for $D>4$,
after symmetry breaking only $SO$(3,1) is visible,
and internal space is static and at the Planck scale.
This explanation is called spontaneous compactification.
Lately, another possible mechanism of compactication 
by branes in string theory
has been discussed.~\cite{witten,riotto}
It is possible
that the Standard Model gauge fields exist on branes 
rather than in the bulk of spacetime.
If this is the case, the situation 
in which gravity exists in the bulk of the 10-dimensional spacetime,
while the Standard Model particles exist on a 3-brane is possible. 
It remains unknown which mechanism of compactification explains 
our universe.
In this paper we consider dynamical compactification only
in the view of multidimensional cosmology.

The simplest multidimensional cosmological model
that is useful for understanding compactification
is the type of extended Friedmann-Robertson-Walker universe.
The pioneering work for this model 
was done by Chodos and Detweiler.~\cite{cd}
They found that the 5-dimensional vacuum Einstein equations
possess the Kasner solution, which describes a universe in which
the internal space shrinks, while the external space expands.
Freund investigated the model in the case of $D=11$
supergravity.~\cite{freund}
Various multidimensional models were investigated,
and the manner in which the existence of internal space might impact 
cosmological issues such as entropy production and inflation
(e.g. Refs.~\cite{ac,ag,sahdev}) was discussed.

On the other hand, the creation of the universe is a 
problem of great interest.
In the standard big-bang cosmology,
the universe appeared from an initial singularity.
When the density is greater than the Planck density,
classical general relativity breaks down
and quantum gravitational effects are large.
Based on quantum gravitational theory,
the scenario of quantum creation of the universe 
was suggested.
Vilenkin asserted that the inflationary universe appears 
through quantum tunneling from nothing.~\cite{vilenkin}
Later Hartle and Hawking suggested the scenario of the creation of the
universe
through a different method.~\cite{hh}
That method is called the Hartle-Hawking no-boundary proposal.
According to this proposal the universe in the quantum era is
Euclidean manifold without ``boundary'',
and through analytic continuation to the Lorentzian manifold,
the classical universe is created.
The condition of analytic continuation gives 
the initial condition of the universe.
Recently, a pre-big-bang scenario has been suggested.~\cite{veneziano}
This scenario is based on string theory and has attractive features. 
However, it has a graceful exit problem.
In this paper we consider quantum cosmology 
from the viewpoint of Hartle and Hawking.

The concept of quantum creation is extended to 
the multidimensional universe with consistency.
Hu and Wu~\cite{wu,hw} discussed the vacuum model
in which the spacetime metric has the form
$R\times S^3\times S^n$ 
in the framework of the Hartle-Hawking no-boundary proposal,
where $R$ is the time, $S^3$ is the external space,
and $S^n$ is the internal space.
They showed that the universe most probably evolves 
with an exponentially expanding external space and a static internal
space.
This solution implies that we observe 4-dimensional spacetime
because internal spaces are compact and static at Planckian scales.
More complicated multidimensional quantum cosmological models 
have been investigated
(e.g. Refs.~\cite{wu1,halliwell1,halliwell2,is,cfp,oy}).

Models based on unified theories, such as supergravity theories,
which are the low energy limits of superstring theories,
would be more realistic 
than the simplest vacuum
model.~\cite{wu1,halliwell2,cfp}
However, it is thought that 
the essence of the creation of the universe
may be included in the vacuum model,
and it is useful for investigating more realistic models. 
Detailed analysis of the vacuum model is, therefore, important.
Hu and Wu discussed only the instanton 
that nucleates the most probable universe, but the instanton solutions 
are unstable with respect to perturbations, and 
the Euclidean classical solutions near the instanton have
effects on the wave function of the universe.
Therefore we analyse instantons in more detail, 
and also investigate dynamical properties of the Euclidean Einstein
equations of
this model by numerical calculations.
As shown in $\S$3, the dynamical properties of the system is of interest,
and it is found that the system has a ``quasi-attractor''.

The plan of the paper is as follows.
In $\S$2, we introduce the multidimensional cosmological model
with cosmological constant ($\Lambda$) but without matter fields.
In $\S$3, we give the constraints
on the signatures of the spatial curvature 
and the cosmological constant
and analyze the Euclidean classical solutions.
In $\S$4, we investigate continuation 
from the Euclidean classical solutions to 
Lorentzian solutions,
and discuss the scenario quantum creation of the universe.
The last section is devoted to conclusions and remarks.
\section{The model}
\label{sec:the Model}

\indent
We consider a $D$($=1+m+n$)-dimensional vacuum universe
with a cosmological constant ($\Lambda$) but without matter fields.
For cosmological purposes, we assume a metric of the form
\begin{equation}
\tilde{g}_{AB}=
\ \left(
\begin{array}{ccc}
-1&0&0\\
0&a^2(t)g_{ij}&0\\
0&0&b^2(t)g_{IJ}\\
\end{array}
\right),\
\end{equation}
where $i$, $j$=1, 2, ..., $m$ and $I$, $J$=1, 2, ..., $n$.
The variables $a$ and $b$ are the scale factors of $m$-dimensional
and $n$-dimensional spaces, respectively.
The gravitational action is described as
\begin{equation}
S=\frac{1}{16\pi \tilde{G}}\int d^{D}x \sqrt{-\tilde{g}}
\biggl[\tilde{R}-2\Lambda\biggr]+S_{boundary},
\end{equation}
where $\tilde{G}$ is the $D$-dimensional gravitational constant,
and $\Lambda$ is the cosmological constant.
The last term $S_{boundary}$ 
is the York-Gibbons-Hawking boundary term.~\cite{york,gh}
Substituting the metric, the action is given by
\begin{eqnarray}
S&=&\frac{1}{16\pi \tilde{G}}\int dt d^m x d^n x
\sqrt{g_m g_n}a^m b^n \nonumber\\
& &\biggl[-m(m-1)\biggl(\frac{\dot{a}}{a}\biggr)^2-2mn\frac{\dot{a}\dot{b}}{ab}
-n(n-1)\biggl(\frac{\dot{b}}{b}\biggr)^2 \nonumber \\
& &+m(m-1)\frac{k_m}{a^2}+n(n-1)\frac{k_n}{b^2}
-2\Lambda \biggr]
,
\end{eqnarray}
where the dots denote derivatives with respect to time $t$,
$g_m$ and $g_n$ are the determinants of the metrics $g_{ij}$ and $g_{IJ}$, 
and $k_m$ and $k_n$ denote the signs of the curvature of external and 
internal space (1, 0 or $-1$), respectively.
The momenta conjugate to the scale factors $a$ and $b$ are given by
\begin{eqnarray}
\pi_a=\frac{\partial{\cal L}}{\partial\dot{a}}
=-\frac{\sqrt{g_i g_e}}{16\pi \tilde{G}} a^m b^n 
\biggl[2m(m-1)\frac{\dot{a}}{a^2}+2mn\frac{\dot{b}}{ab}\biggr]\\
\pi_b=\frac{\partial{\cal L}}{\partial\dot{b}}
=-\frac{\sqrt{g_i g_e}}{16\pi \tilde{G}} a^m b^n 
\biggl[2mn\frac{\dot{a}}{ab}+2n(n-1)\frac{\dot{b}}{b^2}\biggr].
\end{eqnarray}
The Hamiltonian ${\cal H}$ is described as
\begin{eqnarray}
{\cal H}&=&\pi_a \dot{a}+\pi_b \dot{b}-{\cal L}\\
&=&-\frac{\sqrt{g_i g_e}}{16\pi \tilde{G}} a^m b^n 
\biggl[m(m-1)\biggl(\frac{\dot{a}}{a}\biggr)^2+2mn\frac{\dot{a}\dot{b}}{ab}
+n(n-1)\biggl(\frac{\dot{b}}{b}\biggr)^2 +V\biggr].
\end{eqnarray}
 The potential energy $V$ is defined by
\begin{equation}
V=m(m-1)\frac{k_m}{a^2}+n(n-1)\frac{k_n}{b^2}
-2\Lambda.
\end{equation}
The Einstein equations are given by
\begin{eqnarray}
m\frac{\ddot{a}}{a}+n\frac{\ddot{b}}{b}-\alpha=0,\\
\frac{\ddot{a}}{a}+(m-1)\biggl(\frac{\dot{a}}{a}\biggr)^2
+n\frac{\dot{a}\dot{b}}{ab}
+\frac{(m-1)k_m}{a^2}-\alpha=0,\\
\frac{\ddot{b}}{b}+(n-1)\biggl(\frac{\dot{b}}{b}\biggr)^2
+m\frac{\dot{a}\dot{b}}{ab}
+\frac{(n-1)k_n}{b^2}-\alpha=0,
\end{eqnarray}
where
\begin{equation}
\alpha=\frac{2\Lambda}{m+n-1}.
\end{equation}
These equations imply the Hamiltonian constraint
\begin{equation}
m(m-1)\biggl(\frac{\dot{a}}{a}\biggr)^2+2mn\frac{\dot{a}\dot{b}}{ab}
+n(n-1)\biggl(\frac{\dot{b}}{b}\biggr)^2+V=0.
\end{equation}
\section{Euclidean solutions}
\label{sec:Euclidean solutions}
We consider the quantum creation of the universe 
in the framework of the Hartle-Hawking no-boundary proposal.
In the WKB approximation, the Hartle-Hawking wave function
can be expressed in the form
\begin{equation}
\Psi[h_{ij},\Phi] \sim \Sigma_k A_k \exp(-B_k),
\end{equation}
where $B_k$ is the Euclidean action for the solutions 
of the Euclidean field equations, which are compact 
and have the given 3-metric $h_{ij}$ and matter field $\Phi$
on the boundary.
The prefactor $A_k$ represents the fluctuations around these solutions.
Since the factor $\exp(-B_k)$ dominates wave function behavior, 
we analyse the classical solutions of the Euclidean Einstein equations.

The Euclidean Einstein equations are given by
\begin{eqnarray}
m\frac{a''}{a}+n\frac{b''}{b}+\alpha=0,\\
\frac{a''}{a}+(m-1)\biggl(\frac{a'}{a}\biggr)^2+n\frac{a'b'}{ab}
-\frac{(m-1)k_m}{a^2}+\alpha=0,\\
\frac{b''}{b}+(n-1)\biggl(\frac{b'}{b}\biggr)^2+m\frac{a'b'}{ab}
-\frac{(n-1)k_n}{b^2}+\alpha=0,
\end{eqnarray}
where primes denote derivatives with respect to the imaginary time $\tau$.
The Hamiltonian constraint in the Euclidean version is described as
\begin{equation}
m(m-1)\biggl(\frac{a'}{a}\biggr)^2+2mn\frac{a'b'}{ab}
+n(n-1)\biggl(\frac{b'}{b}\biggr)^2-V=0.
\end{equation}

We assume
that $a$$=0$ and/or $b=0$ at $\tau=0$.
The Hartle-Hawking no-boundary proposal gives the boundary condition
for the Euclidean Einstein equations.
According to this proposal,
the Euclidean solutions must be analytic. 
At $\tau=0$, therefore, $a$ and $b$ can be expanded in Taylor series.
This determines the boundary conditions of the Euclidean solutions.
In addition, the external space of the present universe is macroscopic 
and the scale factor of the internal space is 
related to the gauge coupling constant.
Therefore we do not consider trivial solutions, $a(\tau)=0$ and/or
$b(\tau)=0$ for any $\tau$.
These constraints lead to the following boundary conditions:

(a) Boundary Condition 1(BC1)
\begin{eqnarray}
k_m=1,\\
a(0)&=&0,  a'(0)=1,\\
b(0)&=&b_0,  b'(0)=0.
\end{eqnarray}

(b)  Boundary Condition 2(BC2)
\begin{eqnarray}
k_n=1,\\
a(0)&=&a_0,  a'(0)=0,\\
b(0)&=&0,  b'(0)=1.
\end{eqnarray}

The boundary condition BC2 is
the symmetric version of the boundary condition BC1.
These boundary conditions imply that 
either internal space or external space
has positive curvature ($k_m=1$ or $k_n=1$).
We can solve the Euclidean Einstein equations numerically
under these boundary conditions.
Below, the families of Euclidean solutions parameterized by $b_0$ and $a_0$ 
are calculated.

We assume that the Euclidean solutions are connected 
to classical Lorentzian solutions 
at a finite $\tau$, where $a'=b'=0$. 
Due to the Hamiltonian constraint (2.13), 
the value of the potential at the connection surface must be
\begin{equation}
V=\frac{m(m-1)k_m}{a^2}+\frac{n(n-1)k_n}{b^2}
-2\Lambda =0.
\end{equation}

Now we discuss the constraints on 
the signature of the spatial curvatures and the cosmological constant.
First, it is easily shown that either internal space or external
space must have positive curvature ($k_m=1$ or $k_n=1$) from 
the boundary conditions BC1 and BC2.
Second, the condition that the connection surface $V=0$ exists
gives constraints on
the signatures of $k_m$, $k_n$ and $\Lambda$.
For example, when each signature of $k_m$ and $k_n$ is the same,
and $\Lambda=0$, the potential $V(a,b)$ cannot be zero anywhere. 
The combinations of signatures of $k_m$, $k_n$ and $\Lambda$
for which the surface $V=0$ exists, are shown in Table I. 

\begin{table}
 \caption[Possible signatures of $k_m$, $k_n$ and $\Lambda$.]
{The signatures of $k_m$, $k_n$ and $\Lambda$
under the condition that the surface $V=0$ exists.
}
 \label{table}
 \begin{center}
   \begin{tabular}{lccccccccc} \hline \hline
     $(m-1)k_{m}$ & + & + & + & 0 & 0 & 0 & $-$ & $-$ & $-$ \\
     $(n-1)k_{n}$ & + & 0 & $-$ & + & 0 & $-$ & + & 0 & $-$ \\ \hline
     $\Lambda=0$ & no & no & yes  & no 
       & yes & no & yes & no & no\\
     $\Lambda>0$ & yes & yes & yes & yes &  no
       & no & yes & no & no\\
     $\Lambda<0$ & no  & no & yes & no & no
       & yes & yes & yes & yes\\ \hline
   \end{tabular}
 \end{center}
\end{table}

Even if these conditions are satisfied, however, the Euclidean 
solutions cannot be necessarily connected 
to Lorentzian solutions, since in some classes of solutions,
$a$ and $b$ increase monotonically and have no velocity-zero points,
i. e. points for which $\dot{a}=\dot{b}=0$. 
We calculated Euclidean solutions numerically, changing
the initial values, 
and investigated whether these Euclidean solutions can be connected 
to Lorentzian solutions or not.
Except for the case in which 
$(m-1)k_m > 0$, $(n-1)k_n > 0$ and $\Lambda > 0$,
the scale factors $a$ and $b$ vary monotonically,
and the condition $a'=b'=0$ cannot be satisfied at any $\tau$.
Then, we found that
Euclidean solutions can be connected to the Lorentzian region
only for the case 
in which $(m-1)k_m > 0$, $(n-1)k_n > 0$ and $\Lambda > 0$.
After all, 
only the case in which $k_m=k_n=1$ and $\Lambda > 0$ is possible
in our model.
This implies that the internal space cannot be 1-dimensional,
because the curvature term of the Einstein equations is zero 
for 1-dimensional space.
Therefore the dimensions of space $m$ and $n$ must be greater than 2.

We now discuss the Euclidean solutions
in the case that 
$k_m=k_n=1$ and $\Lambda > 0$ in detail.
In this case, there are following three exact solutions:

(1)
\begin{eqnarray}
a&=&\sqrt{\frac{m(m+n-1)}{2\Lambda}}
\sin\biggl(\sqrt{\frac{2\Lambda}{m(m+n-1)}}\tau\biggr),\\
b&=&\biggl[\frac{(n-1)(n+m-1)}{2\Lambda}\biggr]^{1/2}\equiv b_{A}.
\end{eqnarray}

(2)
\begin{eqnarray}
a&=&\sqrt{\frac{(m+n-1)(m-1)}{2\Lambda}}
\sin\biggl(\sqrt{\frac{2\Lambda}{(m+n)(m+n-1)}}\tau\biggr),\\
b&=&\sqrt{\frac{(m+n)(n-1)}{2\Lambda}}
\sin\biggl(\sqrt{\frac{2\Lambda}{(m+n)(m+n-1)}}\tau\biggr).
\end{eqnarray}

(3)
\begin{eqnarray}
a&=&\biggl[\frac{(m-1)(n+m-1)}{2\Lambda}\biggr]^{1/2}\equiv a_{A},\\
b&=&\sqrt{\frac{n(m+n-1)}{2\Lambda}}
\sin\biggl(\sqrt{\frac{2\Lambda}{n(m+n-1)}}\tau\biggr).
\end{eqnarray}
The instanton solutions (1) and (3) were discussed by Hu and Wu.~\cite{wu} 
In addition we have found the instanton solution (2).

\begin{figure}
  \epsfxsize=76mm
  \centerline{\epsfbox{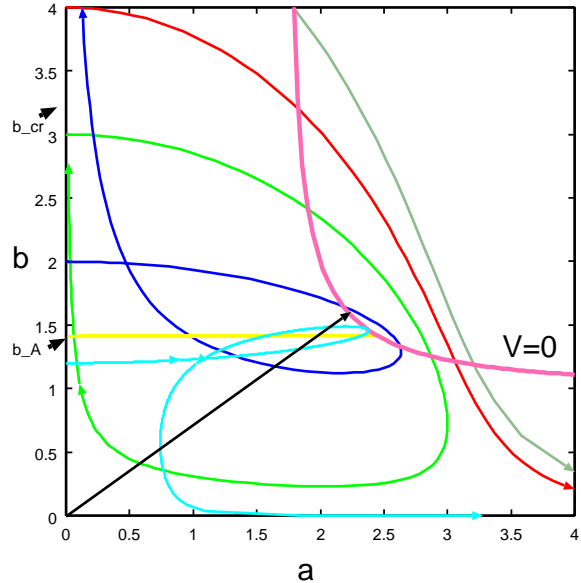}}
\caption{Euclidean solutions in the case
that $m=3$, $n=2$, $k_m=k_n=1$, $\Lambda=1$
and with the boundary condition BC1 (Eqs.(3.3) and (3.4)).
The critical value $b_{cr}$ is 3.16.}
\label{fig:1}
\end{figure}

\begin{figure}
  \epsfxsize=76mm
  \centerline{\epsfbox{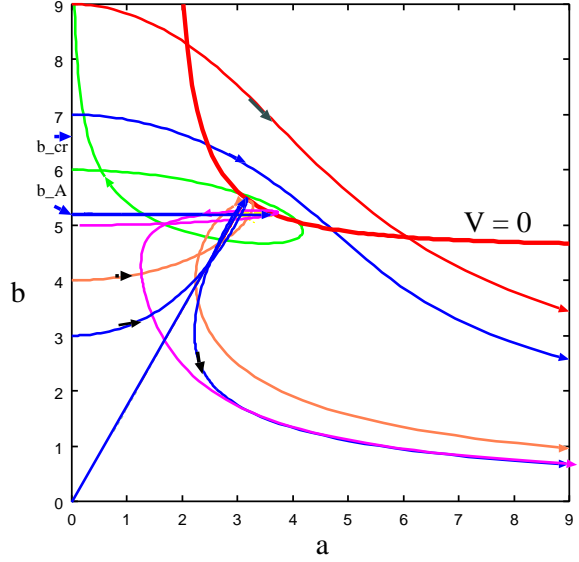}}
\caption{Euclidean solutions in the case
that $m=3$, $n=7$, $k_m=k_n=1$, $\Lambda=1$
and with the boundary condition BC1 (Eqs.(3.3) and (3.4)).
The critical value $b_{cr}$ is 6.71.}
\label{fig:2}
\end{figure}

\begin{figure}
  \epsfxsize=76mm
  \centerline{\epsfbox{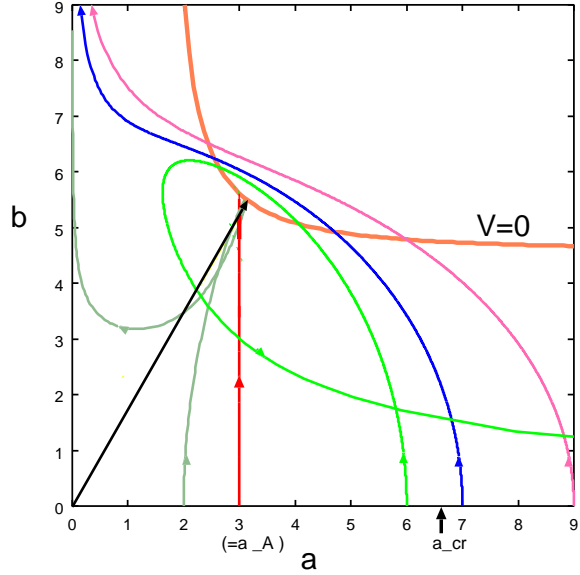}}
\caption{Euclidean solutions in the case
that $m=3$, $n=7$, $k_m=k_n=1$, $\Lambda=1$
and with the boundary condition BC2 (Eqs.(3.6) and (3.7)).
The critical value $a_{cr}$ is 6.71.}
\label{fig:3}
\end{figure}

In Figs. 1$\sim$3, the paths in the $a-b$ plane are shown.
Here we set $\Lambda =1$ since we can always set $\Lambda=1$
by rescaling the cosmic time.
Figs. 1 and 2 are for the case of the boundary condition
BC1 (Eqs. (3.3) and (3.4)).
Fig. 3 is for the case of 
the boundary condition BC2 (Eqs. (3.6) and (3.7)).
The essential property of the dynamical system 
is independent of the dimension of the $g_{ij}$ or $g_{IJ}$ spaces,
as shown in Figs. 1 and 2. 
Note that this dynamical system has a characteristic point $A$:
\begin{equation}
(a_A,b_A)
=\biggl(\biggl[\frac{(m-1)(n+m-1)}{2\Lambda}\biggr]^{1/2},
\biggl[\frac{(n-1)(n+m-1)}{2\Lambda}\biggr]
^{1/2}\biggr).
\end{equation}
The trajectories of the Euclidean solutions 
on the scale factor plane $a$-$b$ 
can be classified into following three classes.
First, if $b_0$ (or $a_0$) is greater than
a critical value $b_{cr}$ (or $a_{cr}$),
\begin{equation}
b_{cr}, a_{cr}=\sqrt{\frac{(D-1)(D-2)}{2\Lambda}},
\end{equation}
the trajectories flow to $(a,b)\rightarrow (\infty, 0)$
(or (0,$\infty$)) directly.
Second, if $b_0$ (or $a_0$) is less than a critical value $b_{cr}$ 
(or $a_{cr}$)
but greater than $b_{A}$ (or $a_{A}$),
the trajectories go around the point $(a_A,b_A)$
and flow to $(a,b)\rightarrow (0, \infty)$ (or ($\infty$,0)).
When $b_0$ (or $a_0$) is just equal to $b_{A}$ (or $a_{A}$),
$b$ (or $a$) is constant (the solution (1) (or (3)))
and the trajectory is a straight line.
Third, if $b_0$ (or $a_0$) is lower than $b_{A}$ (or $a_{A}$),
the trajectories go around the point $ (a_A,b_A)$
and flow to $(a,b)\rightarrow (\infty,0)$ (or (0,$\infty$)).

Note that the critical values $b_{cr}$ and $a_{cr}$ depend 
only on the dimension of the spacetime $D$ as described by (3.20).
We call the point $(a_A,b_A)$, around which the trajectories circulate,
the ``quasi-attractor.'' 
The ``quasi-attractor'' plays a role
in gathering 
the trajectories of the classical Euclidean solutions
to the instanton solutions (Eqs. (3.9)-(3.14)),
and approximately connecting
to the Lorentzian solutions.
In the next section we discuss the properties of the Lorentzian solutions.
\section{Lorentzian solutions}
\label{sec:Lorentzian solutions}
In this section we investigate the Lorentzian solutions 
that are continued from
the Euclidean solutions we have presented in the preceding section 
in the case that the curvatures of both external and internal spaces are
positive and $\Lambda > 0$.
The Euclidean manifold is connected to the Lorentzian manifold,
where the velocities of the scale factors are vanishing,
and it can be shown that
at the connection surface, the value of the potential $V$ 
must be vanishing.

There are three following exact solutions which are 
analytically continued from instantons:

(1)
\begin{eqnarray}
a&=&\sqrt{\frac{m(m+n-1)}{2\Lambda}}
\cosh\biggl(\sqrt{\frac{2\Lambda}{m(m+n-1)}}t\biggr),\\
b&=&\biggl[\frac{(n-1)(n+m-1)}{2\Lambda}\biggr]^{1/2}.
\end{eqnarray}

(2)
\begin{eqnarray}
a&=&\sqrt{\frac{(m+n)(m-1)}{2\Lambda}}
\cosh\biggl(\sqrt{\frac{2\Lambda}{(m+n)(m+n-1)}}t\biggr),\\
b&=&\sqrt{\frac{(m+n)(n-1)}{2\Lambda}}
\cosh\biggl(\sqrt{\frac{2\Lambda}{(m+n)(m+n-1)}}t\biggr).
\end{eqnarray}

(3)
\begin{eqnarray}
a&=&\bigl[\frac{(m-1)(n+m-1)}{2\Lambda}\biggr]^{1/2},\\
b&=&\sqrt{\frac{n(m+n-1)}{2\Lambda}}
\cosh\biggl(\sqrt{\frac{2\Lambda}{n(m+n-1)}}t\biggr).
\end{eqnarray}

In the preceding section we found that the ``quasi-attractor'' plays
a role in gathering the paths of
the classical Euclidean solutions to the instanton
solutions (Eqs. (3.9) $\sim$(3.14)), and approximately connecting
to Lorentzian solutions.
Strictly speaking, 
the Euclidean solutions near these instanton solutions 
cannot analytically continue to Lorentzian solutions.
Since these nearby solutions exists very densely, 
their effects are necessary 
for estimating the wave function of the universe.
Therefore we make detailed analysis of these solutions.
The Lorentzian solutions under the initial conditions 
that the velocities of the scale factors on the connection
surface $V(a,b)=0$ are zero are approximately continued 
from these solutions.
Thus we calculate the Lorentzian solutions continued approximately 
from these solutions. 
In Fig. 4 the Lorentzian solutions in the $a$-$b$ plane are shown.
The solutions between the solutions (1) and (3) 
show that both external and internal spaces expand monotonically.
On the other hand, the outer solutions of the exact solution (3) 
show that internal space expands 
and external space contracts.
The outer solutions of the exact solution (1), 
which are nothing but the symmetric version of the former solutions,
show that external space expands 
and the internal space contracts (Fig. 5).
The last solutions imply that
external space of the observed universe is macroscopic
and internal space of the observed universe is microscopic.

\begin{figure}
  \epsfxsize=76mm
  \centerline{\epsfbox{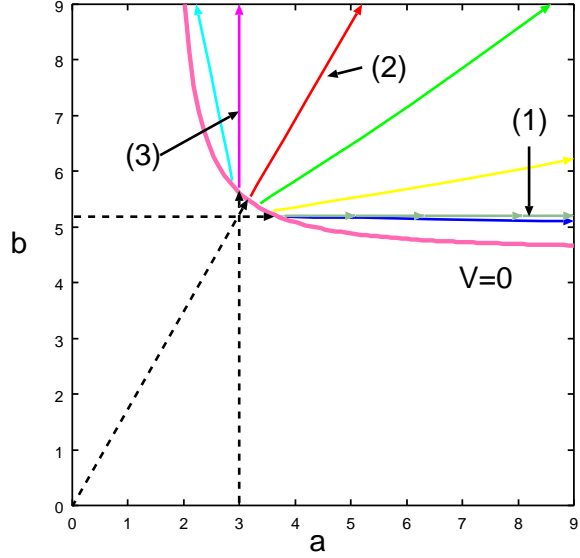}}
\caption{Lorentzian solutions 
in the case $m=3$, $n=7$, $k_m=k_n=1$ and $\Lambda=1$.
The number in this figure represents that of the Lorentzian analytic
solutions
(Eqs. (4.2)$\sim$(4.7)).
The dotted lines represent the instantons
(Eqs. (3.9)$\sim$(3.14))}
\label{fig:4}
\end{figure}

\begin{figure}
  \epsfxsize=76mm
  \centerline{\epsfbox{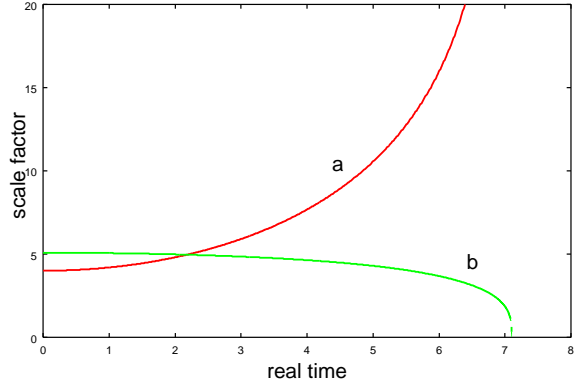}}
\caption{Lorentzian solution 
in the case $m=3$, $n=7$, $k_m=k_n=1$, $\Lambda=1$                 
and the initial condition are 
$a=4$, $\dot{a}=0$, $b=[21/(1-3/4^2)]^{1/2}$, $\dot{b}=0$}
\label{fig:5}
\end{figure}
\section{Summary and discussion}
\label{sec:summary and discussion}
In this paper we have investigated 
the quantum creation of a multidimensional universe
with a cosmological constant but without matter fields
in the framework of the Hartle-Hawking no-boundary proposal.
The most interesting result is that 
there are ``quasi-attractor,'' i.e., 
most trajectories of the classical solutions of the Euclidean Einstein
equation go around on the $a$-$b$ plane, 
independently of the initial values, 
provided that the initial values
of the  scale factors are smaller than a critical value
in the case that 
the curvatures of both the external and internal spaces are positive
and $\Lambda>0$. 
This characteristic
behavior of the evolution of the scale factors in Euclidean time is
essentially independent of the numbers of external and internal spaces
dimensions, provided that $m, n \ge 2$.

Since the trajectories approach the quasi-attractor at the first
step, it plays a role in gathering the paths of the classical
Euclidean solutions to the instantons ((3.9)
$\sim$(3.14)) independently of the initial values.
Then, it is very natural
to assume that  the wave function has a hump 
near this quasi-attractor point. 
There Lorentzian solutions which start from the points on
the continuation surface near the quasi-attractor give a dominant
contribution to the wave function in the Lorentzian region.
As the example is in Fig. 5 shown, there are solutions
for which we can interpret
that dynamical compactification takes place;
i.e. the external space evolves to the macroscopic realm 
and the internal space remains microscopic.

In order to understand the role of this quasi-attractor in the wave
function and the creation of the universe more deeply,
and to confirm the above conjecture, 
it is necessary to analyze the Wheeler-De Witt equations   
and to calculate the wave function of the universe.
The Wheeler-De Witt equations of the multi-dimensional vacuum universe
have been analysed by Chmielowski,~\cite{c} but only for models
with zero cosmological constant. 
According to his result, if neither of the subspaces has negative
curvature,
then there exists no Lorentzian solution.
Obviously, interesting cases are those of the positive cosmological
constant,
as we have discussed in the present paper. 
Analysis of the Wheeler-De Witt 
equations in this model is under progress.

  In order to make more realistic models, we must consider more
complicated models on the basis of fundamental theories.
As shown in Fig. 5, the solutions which show
that the external space expands
and the internal spaces collapses in finite time are adequate models for
our universe, provided that the collapse of inner space is halted.
Furthermore, because the volume of the internal space is related to
the constants of nature, 
the present internal space must be stable at the order of the Planck scale.
Fortunately, it is considered that the collapse of the internal space
is halted
by the pressure of the gauge fields or by a quantum effects,
and the internal space is stabilized.
It was shown that in $D=6$, $N=2$ supergravity theories, 
the solution of (the 4-dimensional Friedmann universe)$\times$
(a constant $S^2$) is the attractor; i.e.,
all cosmological solutions starting from arbitrary initial conditions
approach 
to the above spacetime asymptotically.~\cite{mn}
Moreover, stable compactification of the extra dimensions
by quantum effects 
was studied. 
The quantum corrections to the effective potential are
attributed to the Casimir effects in many works
(e.g. Refs.~\cite{gz,oy} and references therein).
In Ezawa et al,~\cite{ezawa} 
the quantum effects 
of higher curvature gravity theories are investigated.
From these investigations, if is known that 
an effective potential of the model obtained
under dimensional reduction to a 4-dimensional effective theory
has minima at the Planck scale for the scale factors of internal spaces.
Therefore the internal spaces can be stable.

Our work is simply based on canonical quantum gravity in general
relativity.
It is, however, thought that general relativity is a low-energy limit
of the ultimate unified theory.
Although the Einstein equation would be necessarily modified at high
energy, it must include essential points of the ultimate theory.
It is natural to conjecture that the ultimate theory may include
the present simple model as an approximate description.
It seems that analysis of the simple model given in the present paper
may be useful for further investigations of the basis of fundamental
theories such as superstring theories and supergravity theories. 
\section*{Acknowledgements}
We are very grateful to Dr. Takeshi Chiba and Dr. Keita Ikumi 
for useful discussions. 
One of the authors (K. S. ) thanks Neil Turok
for stimulating discussions.
This work was supported in part 
by the grant-in-Aid for Scientific Research (07CE2002) 
of the Ministry of Education, Science, Sports and Culture in Japan.

\end{document}